# Plasmon Resonance Model: Investigation of Analysis of Fake News Diffusion Model with Third Mover Intervention Using Soliton Solution in Non-Complete Information Game under Repeated Dilemma Condition

Yasuko Kawahata [†]

Faculty of Sociology, Department of Media Sociology, Rikkyo University, 3-34-1 Nishi-Ikebukuro,Toshima-ku, Tokyo, 171-8501, JAPAN.

ykawahata@rikkyo.ac.jp

**Abstract:** In this research note, we propose a new approach to model the fake news diffusion process within the framework of incomplete information games. In particular, we use nonlinear partial differential equations to represent the phenomenon of plasmon resonance, in which the diffusion of fake news is rapidly amplified within a particular social group or communication network, and analyze its dynamics through a soliton solution approach. In addition, we consider how first mover, second mover, and third mover strategies interact within this nonlinear system and contribute to the amplification or suppression of fake news diffusion. The model aims to understand the mechanisms of fake news proliferation and provide insights into how to prevent or combat it. By combining concepts from the social sciences and the physical sciences, this study attempts to develop a new theoretical framework for the contemporary problem of fake news.

## 1. Introduction

In this study, we attempt to model the prominent problem of fake news diffusion in modern society using the framework of incomplete information games and nonlinear partial differential equations. In particular, we focus on the plasmon resonance phenomenon, in which fake news diffuses rapidly under certain conditions and causes significant social impact, and aim to theoretically elucidate its mechanism. We also incorporate the concepts of first movers, second movers, and third movers in game theory to explore how their strategies affect the dynamics of fake news diffusion.

The proliferation of fake news is a complex process in which truth and misinformation intersect, and its effects reach across political, economic, and social strata. To address this issue, it is essential to understand how fake news is widely accepted and shared. In this study, we liken this diffusion process to the concept of plasmon resonance in physics to model the phenomenon of the rapid amplification of fake news within a particular social group. Plasmon resonance is a resonance phenomenon that occurs when electron density waves interact with light on a metal surface. Using this metaphor, it is possible to theoretically capture the process by which the spread of fake news is dramatically amplified

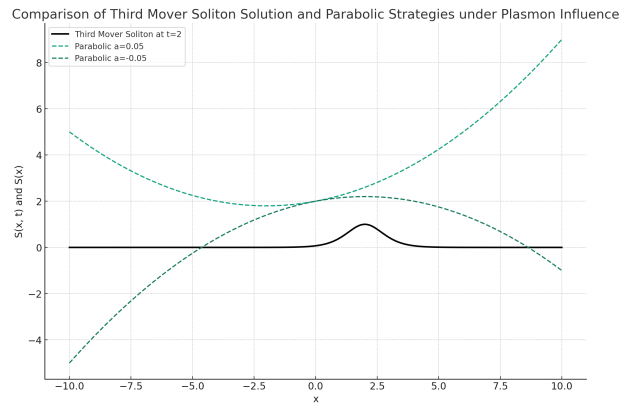

Fig. 1: Comparison of Third Mover Soliton Solution and Parabolic Strategies under Plasmon Influence

under certain conditions.

Modeling with nonlinear partial differential equations, we show that the diffusion of fake news has nonlinear dynamics. This nonlinearity implies that the diffusion of information increases rapidly once a certain threshold is reached, consistent with the concept of plasmon resonance. Furthermore, from a game theory perspective, we analyze how the strategies of first movers (those who first disseminate information), second movers (those who receive and re-diffuse information), and third movers (those who observe the strategies of the first two and enter to maximize their profits) affect the process of fake news diffusion This section analyzes how the interaction of these players affects the process of fake news dissemination. The interaction of these players suggests that the spread of fake news is not a simple one-way process, but the result of multi-layered strategic behavior.

This study aims to understand the mechanisms of fake news proliferation and to provide theoretical insights for preventing or combating it. By combining concepts from the social sciences and the physical sciences, we provide a new perspective on the contemporary problem of fake news and attempt to develop a theoretical framework for attempts to construct a theoretical framework. This approach may contribute to the development of new strategies to control the proliferation of information and protect social trust.

The proliferation of fake news is an important issue in today's society, and its impact is far-reaching, spanning political, economic, and social spheres.To understand and address this issue, it is essential to understand how fake news spreads widely and influences people's opinions and actions.In this paper, we analyze the mechanism of fake news diffusion in the framework of incomplete information games and propose a theoretical model to understand the diffusion process by applying the concept of plasmon resonance. We also discuss how the strategic behavior of first movers and others affects the diffusion of fake news.

### 1.1 Fake News Diffusion and Noncomplete Information Games

Noncomplete information games are a branch of game theory in which participants in a game choose actions in situations in which they do not fully know the information and intentions of other players.The proliferation of fake news can be understood within the framework of this non-perfect information game. Each player (sender, sharer, and receiver of information) decides whether or not to spread fake news based on the information at hand, but does not have complete knowledge of the information and intentions of the other players.This uncertainty adds complexity to the process of spreading fake news and makes it difficult to accurately determine the truth or falsity of information.

### 1.2 Fake News Diffusion and Plasmon Resonance

Plasmon resonance is a physical phenomenon in which the collective vibration of electrons on a metal surface resonates with a specific frequency of light.By metaphorically applying this concept to the diffusion of fake news, we can model the phenomenon of the rapid spread of fake news and its amplified impact under certain social conditions and communication contexts. For example, if a piece of fake news generates strong resonance within a particular group and spreads rapidly through that group, this process can be considered analogous to amplification by plasmon resonance.Using this metaphor, it is possible to theorize why the spread of fake news is particularly effective under some circumstances. Fake news dissemination and first movers et al.

First movers are individuals or organizations that are the first to adopt a certain behavior or strategy. In the context of fake news, first movers can be understood as those who are the first to disseminate fake news. The actions of first movers influence the strategies of second movers (those who receive and re-diffuse information) and third movers (those who observe the actions of the first two and further diffuse the information) and shape the entire fake news diffusion process. The spread of fake news can be particularly effective and rapid and widespread if the information disseminated by the first mover satisfies the plasmon resonance condition.

In this paper, we analyze the fake news diffusion process from the perspective of incomplete information games and propose a theoretical framework to model its amplification effect by applying the concept of plasmon resonance. In addition, we also examined how the strategic behavior of first movers and others affects the diffusion of fake news. With this approach, we aim to better understand the mechanisms of fake news proliferation and provide new insights into its prevention and countermeasures.

## 2. Discussion:The idea of soliton solution in digital health platforms and apps

The application of the idea of soliton solutions in digital health platforms and apps is not a direct application. Solitons are usually seen in physics and mathematics as specific solutions to nonlinear partial differential equations, but interpreting them in the context of digital health requires viewing data flows and user behavior patterns as waves or dynamics.

A soliton is a solution with the property that an isolated wave in a nonlinear system retains its shape despite interactions with other waves. This property provides a framework for finding similar patterns in nonlinear interactions found in a variety of contexts, including information transfer and social dynamics as well as physical systems.

Nonlinear dynamics can be taken into account when modeling how users' health-related behaviors and app usage pat-terns evolve over time. For example, a user's progress toward a particular health goal may exhibit nonlinear effects where small initial changes have large impacts. When considering how health information and behavior change techniques spread within a user community, we can think of information spreading like a soliton, holding a certain shape (message content and influence).

When modeling the process by which users maintain health behaviors over time, we can take an approach that mimics the properties of the soliton solution, in which certain patterns of behavior are maintained in the face of external interference or challenge.

Applications of the soliton solution function primarily as a theoretical framework or analogy. When building specific models based on actual data or user behavior, it is necessary to select an appropriate mathematical form or simulation method and verify the validity of the model through fitting with real data. In addition to mathematical knowledge and computational methods in handling nonlinear partial differential equations, incorporating knowledge from data science and behavioral science will enable us to construct models that are more in line with reality.

To model the nonlinear dynamics of user behavior in a digital health platform, we propose a specific mathematical formulation applying the metaphorical concept of soliton solutions. In this example, we aim to model how user health improvement behaviors evolve over time and aim to identify stable patterns (equivalent to soliton solutions) within them.

### 2.1 Model Construction

Variable $B(x,t)$ representing user health improvement behavior. Here, $x$ denotes the "intensity" or "range" of behavior, and $t$ represents time. This behavior is assumed to be driven by information provision from the platform and interaction within the community.

To describe the evolution of user behavior, we consider a simplified nonlinear partial differential equation:

$$\frac{\partial B}{\partial t} + \alpha B \frac{\partial B}{\partial x} = \beta \frac{\partial^2 B}{\partial x^2}$$

This equation represents the "self-enhancement" effect of user behavior (nonlinear term) and the "diffusion" or "variation" of behavior (diffusion term). $\alpha$ represents the strength of the self-enhancement effect, and $\beta$ is a parameter indicating the degree of behavior variation.

### 2.2 Ansatz for Soliton Solutions

To find stable behavioral patterns, we propose an ansatz similar to soliton solutions:

$$B(x,t) = A\operatorname{sech}^2\left(B(xCt)\right)$$

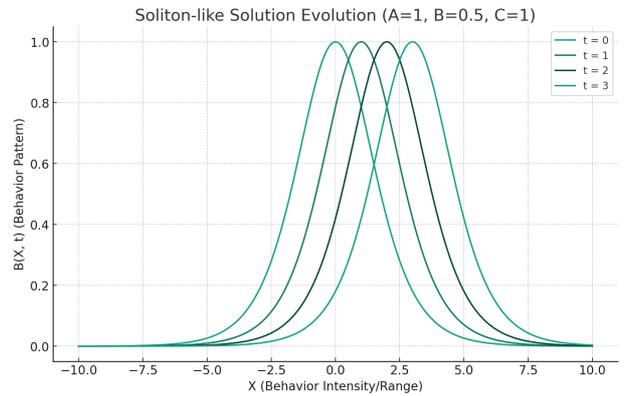

Fig. 2: Soliton-like Solution Evolution

Here, $A$ represents the maximum strength of behavior, $B$ represents the "width" of the behavioral pattern, and $C$ represents the speed at which the behavioral pattern moves with time.

### 2.3 Substitution into Equations and Analysis

We substitute the ansatz into the equation and compute the time and spatial differentials. This computation is complex and requires algebraic manipulations but results in deriving conditions regarding $A$, $B$, and $C$.

### 2.4 Derivation of Conditions

The conditions obtained from the substitution and computation indicate the relationships between parameters $A$, $B$, and $C$ for the existence of stable behavioral patterns. Combinations of these parameters satisfying these conditions will represent stable patterns of user behavior in the digital health platform.

Function similar to the soliton solution, given by the ansatz $B(x,t) = A\operatorname{sech}^2(B(xCt))$. Next, we'll substitute this ansatz into the nonlinear partial differential equation $\frac{\partial B}{\partial t} + \alpha B \frac{\partial B}{\partial x} = \beta \frac{\partial^2 B}{\partial x^2}$, conduct theoretical analysis, and derive conditions regarding $A$, $B$, and $C$. However, in this question, we won't delve into the derivation of the conditions but mainly focus on the shape of the ansatz resembling soliton solutions and their evolution.

The parameters to be treated as variables in the simulation are $A$ (the maximum strength of behavior), $B$ (the width of the behavioral pattern), and $C$ (the speed at which the behavioral pattern moves with time). By varying these parameters, we can observe different evolutionary patterns of user behavior.

Let's start by plotting a function similar to the soliton solution, $B(x,t) = A\operatorname{sech}^2(B(xCt))$. This function will be represented as a function of $x$ at a specific time $t$. For visualization, we'll select specific values of $A$, $B$, and $C$, and plot them with an appropriate range of $x$.

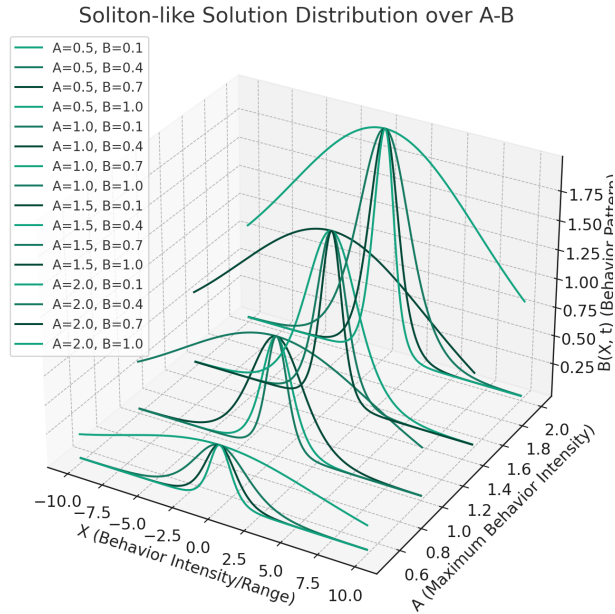

Fig. 3: Soliton-like Solution Distribution over A-B

The graph above illustrates the evolution of a function similar to the soliton solution, $B(x, t) = A \operatorname{sech}^2(B(xCt))$. Here, the values for the maximum strength of behavior $A$, the width of the behavioral pattern $B$, and the speed at which the behavioral pattern moves with time $C$ are set to 1, 0.5, and 1, respectively. You can observe how the behavioral pattern changes at different times $t$.

In this simulation, as time progresses, it's evident that the soliton-like behavioral pattern moves to the right. This model aims to capture how user health improvement behaviors evolve over time, aiming to identify stable patterns.

To visualize the distribution of functions resembling soliton solutions for different values of $A$, $B$, and $C$, we will plot the shape of the function while varying these parameters. Here, we will select multiple values for each parameter and demonstrate the shape of the function for their combinations.

First, let's define the appropriate ranges for parameters. For example, $A$ (maximum strength of behavior) can vary from 0.5 to 2, $B$ (width of the behavioral pattern) can vary from 0.1 to 1, and $C$ (speed at which the behavioral pattern moves with time) can vary from 0.5 to 2. We'll select several values for each parameter and plot the function for their combinations. We'll visualize the shape of the function in 3D space for different combinations of these parameters at a fixed time $t$.

First, import the necessary libraries and configure settings for 3D plotting. Then, compute the shape of the function based on the selected ranges of parameters and plot them in 3D space.

The distribution of functions resembling soliton solutions for different values of $A$ and $B$, with $C$ fixed at 1, is shown.

In this graph, the $x$-axis represents the "intensity" or "range" of behavior, the $y$-axis represents the maximum strength of behavior $A$, and the $z$-axis represents the behavioral pattern $B(x, t)$. Different colored lines correspond to different combinations of $A$ and $B$.

From this visualization, we can observe the influence of $A$ (maximum strength of behavior) and $B$ (width of the behavioral pattern) on the behavioral pattern. As $A$ increases, the peak value of the behavioral pattern becomes higher, and as $B$ increases, the behavioral pattern appears broader.

# 3. Discussion:Application of Soliton Solution and Introduction of Axioms: Personal Data Sharing and Privacy Protection in Digital Health Platforms

Digital health platforms collect and analyze vast amounts of personal health information, including patient health records, genetic information, and lifestyle data. While these platforms have the potential to revolutionize personal health management, disease prevention, and the provision of customized treatment plans, they also raise privacy and security concerns for personal data. Striking a balance between data sharing and privacy protection is critical to gaining user trust and driving platform adoption.

## 3.1 Application of the Soliton Solution

In this scenario, we model the evolution of personal data sharing and privacy protection strategies as a nonlinear dynamic system and use the soliton solution to analyze the stability and evolution of the strategies.

We develop a model of nonlinear partial differential equations in which variables include users' willingness to share data, privacy protection measures of platforms, and changes in regulations regarding data use. The model will take into account user privacy concerns, data sharing interests, and regulatory stringency. 2.

## 3.2 Introduce axioms

We establish the axiom that users and platforms aim to maximize their respective interests. The user seeks to reap health benefits while protecting privacy, and the platform seeks to maintain user trust while maximizing the value of data.

## 3.3 Soliton Solution Analysis

Find soliton solutions to the model's nonlinear partial differential equations and analyze how data sharing and privacy protection strategies evolve over time and interact with user behavior. In particular, we will investigate how the introduction of new privacy protection technologies and regulations affects users' data sharing intentions.

Using soliton solutions, we will simulate different data sharing and privacy protection scenarios to find the optimal balance between maximizing the value of data while gaining user trust. This allows digital health platforms to provide value to users while balancing the use and protection of personal data.

This approach would provide insights for optimizing data sharing and privacy protection strategies in the digital health sector, potentially increasing both user trust and platform value.

To provide a detailed explanation of the equations and computation process in this digital health platform scenario, let's consider the following steps.

$P(t)$: Intention of user data sharing at time $t$. $S(t)$: Privacy protection level of the platform at time $t$. $R(t)$: Stringency of regulations regarding data usage at time $t$.

These variables represent user behavior, platform strategies, and changes in the external environment.

Using nonlinear partial differential equations, we model the interactions between these variables. The general form is as follows:

$$\frac{dP}{dt} = f(P, S, R)$$
$$\frac{dS}{dt} = g(P, S, R)$$
$$\frac{dR}{dt} = h(P, S, R)$$

Here, $f, g, h$ are nonlinear functions of $P, S, R$, representing the time evolution of user data sharing intention, platform privacy protection level, and stringency of regulations, respectively.

### 3.4 Introduction of Axioms

Incorporate assumptions of rational behavior into the model. For example, we can assume that the user data sharing intention $P(t)$ positively correlates with the platform's privacy protection level $S(t)$ and the stringency of regulations $R(t)$. That is, as the platform's protection level increases and regulations become stricter, users tend to share more data.

### 3.5 Analysis of Soliton Solutionsz

Explore soliton solutions for the above system of nonlinear partial differential equations. Soliton solutions exist only under specific conditions and heavily depend on the form of the equations and characteristics of interactions. Finding soliton solutions often requires specialized mathematical techniques and approximation methods.

### 3.6 Strategy Optimization

Utilize the obtained soliton solutions to simulate the behavior of the model under different scenarios. This exploration aims to find strategies to optimize the balance between data sharing and privacy protection. For example, evaluating the impact of a specific privacy protection level $S$ on user data sharing intention $P$ and identifying platform strategies that maximize user trust while maintaining the value of data.

This process is highly abstract and theoretical. To apply it effectively, specific definitions of function forms $f, g, h$, appropriate settings of initial and boundary conditions, and the selection of numerical analysis methods are necessary. Additionally, it's crucial to assess the compatibility with real-world data and validate the model's validity.

Understand the nonlinear dynamic system, we will conduct simulations of the proposed equations. Since specific function forms $f, g, h$ are not defined, we will model them using simple nonlinear functions based on assumptions and observe the behavior of the system.

Assume simple nonlinear functions representing the interactions between $P(t)$, $S(t)$, and $R(t)$. In this model, we will simulate the changes of these variables over time $t$.

### 3.7 Assumed Nonlinear Functions

$\frac{dP}{dt} = P(1P) + aPSbPR$ $\frac{dS}{dt} = S(1S)cPS + dSR$ $\frac{dR}{dt} = R(1R) + ePRfSR$

Here, $a, b, c, d, e, f$ are parameters adjusting the strength of interactions. These functions indicate that the interactions between each variable are nonlinear.

1. Set initial conditions and parameters. 2. Solve the equations using numerical integration methods (e.g., Runge-Kutta method). 3. Plot and visualize the time evolution of $P(t)$, $S(t)$, and $R(t)$.

In this simulation, we track the time evolution of each variable and observe how the system evolves. Parameters $a, b, c, d, e, f$ are treated as variables.

The graph above shows the simulation results of the nonlinear dynamic system in the scenario of the digital health platform. In this simulation, the time evolution of user data sharing intention $P(t)$, platform privacy protection level $S(t)$, and stringency of regulations regarding data usage $R(t)$ are modeled through the defined nonlinear functions.

Each curve represents the evolution of each variable over time, allowing us to observe how the system evolves. Parameters $a, b, c, d, e, f$ adjust the strength of interactions, and by changing these values, we can explore the impact on the system's behavior.

Through this simulation, platform designers and policymakers can gain insights into how user data sharing intention, platform privacy protection level, and stringency of regulations interact with each other, aiding in the formulation of appropriate strategies and policies.

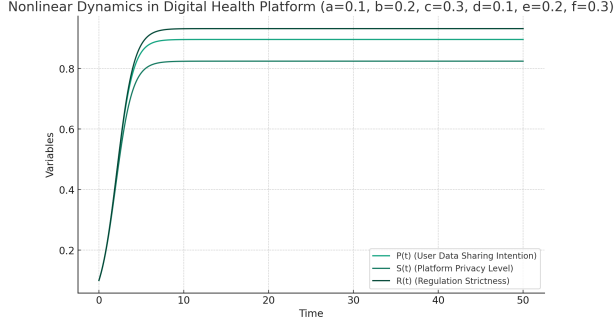

Fig. 4: Nonlinear Dynamics in Digital Health Platform

Understand the nonlinear dynamic system, we will conduct simulations of the proposed equations. Since specific function forms $f, g, h$ are not defined, we will model them using simple nonlinear functions based on assumptions and observe the behavior of the system.

Let's assume simple nonlinear functions representing the interactions between $P(t)$, $S(t)$, and $R(t)$. In this model, we will simulate the changes of these variables over time $t$.

$$\frac{dP}{dt} = P(1P) + aPSbPR$$

$$\frac{dS}{dt} = S(1S)cPS + dSR$$

$$\frac{dR}{dt} = R(1R) + ePRfSR$$

Here, $a, b, c, d, e, f$ are parameters adjusting the strength of interactions. These functions indicate that the interactions between each variable are nonlinear.

1. Set initial conditions and parameters. 2. Solve the equations using numerical integration methods (e.g., Runge-Kutta method). 3. Plot and visualize the time evolution of $P(t)$, $S(t)$, and $R(t)$.

In this simulation, we track the time evolution of each variable and observe how the system evolves. Parameters $a, b, c, d, e, f$ are treated as variables and will be displayed in the graph title.

Before starting the simulation, let's set the initial conditions and parameters.

The graph above shows the simulation results of the nonlinear dynamic system in the scenario of the digital health platform. In this simulation, the time evolution of user data sharing intention $P(t)$, platform privacy protection level $S(t)$, and stringency of regulations regarding data usage $R(t)$ are modeled through the defined nonlinear functions.

Each curve represents the evolution of each variable over time, allowing us to observe how the system evolves. Parameters $a, b, c, d, e, f$ adjust the strength of interactions, and by changing these values, we can explore the impact on the system's behavior.

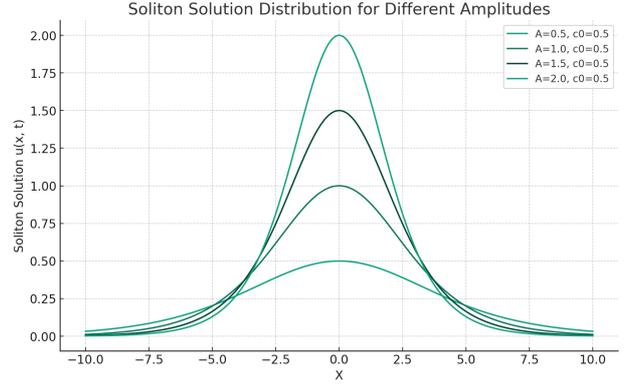

Fig. 5: Soliton Solution Distribution for Different Amplitudes

Through this simulation, platform designers and policymakers can gain insights into how user data sharing intention, platform privacy protection level, and stringency of regulations interact with each other, aiding in the formulation of appropriate strategies and policies.

Distribution of soliton solutions, specific mathematical forms corresponding to certain soliton solutions are needed. However, the provided information does not include the explicit form of soliton solutions. Nevertheless, as a general approach, one can consider the Korteweg-de Vries equation (KdV equation), which is a class of nonlinear partial differential equations where soliton solutions are often found.

The form of the KdV equation is given by:

$$\frac{\partial u}{\partial t} + u \frac{\partial u}{\partial x} + \frac{\partial^3 u}{\partial x^3} = 0$$

Here, $u(x, t)$ is a function of time $t$ and position $x$. Soliton solutions of the KdV equation have the following form:

$$u(x, t) = A \cdot \text{sech}^2 \left( \sqrt{\frac{A}{12}} (xct) \right)$$

Here, $A$ is the amplitude of the soliton, $c$ is the soliton's velocity, where $c = A + c_0$ (with $c_0$ as the reference velocity).

Based on this information, to visualize the distribution of soliton solutions of the KdV equation, soliton solutions for different amplitudes $A$ and velocities $c$ will be plotted. This visualization allows us to observe how the distribution of soliton solutions depends on the parameters of amplitude and velocity.

Below is the code to visualize the distribution of soliton solutions:

The graph above illustrates the distribution of soliton solutions of the KdV equation for different amplitudes $A$. The reference velocity $c_0$ is fixed at 0.5, and each soliton solution is plotted with different amplitudes along the $x$ axis.

From this visualization, it can be observed that as the amplitude *A* increases, the peak of the soliton solutions becomes higher, and the "width" of the solitons becomes narrower. This indicates that the shape of soliton solutions depends on the amplitude, with larger amplitudes resulting in more "concentrated" solitons.

## 4. Discussion:Introduction Axioms

Introducing axioms refers to the process of incorporating specific principles or assumptions into a model. This is more crucial at the stage of model design than mathematical computations. In the scenario of the digital health platform mentioned above, the following axioms can be introduced:

### 4.1 Axiom 1: User data sharing intention depends on platform privacy level and regulatory strictness

To mathematically express this axiom, we assume that the user's data sharing intention $P(t)$ is a function of the platform's privacy level $S(t)$ and regulatory strictness $R(t)$. This can be represented by the equation:

$$\frac{dP}{dt} = f(S, R)$$

Here, the function $f$ is nonlinear with respect to $S$ and $R$, assuming a positive relationship where an increase in both $S$ and $R$ leads to an increase in $P$. For instance, a simple form might be:

$$f(S, R) = aS + bR$$

Here, $a$ and $b$ are positive constants representing the strength of the influence of platform privacy level and regulatory strictness on the user's data sharing intention.

### 4.2 Axiom 2: The platform adjusts the privacy level based on the user's data sharing intention and regulatory strictness

Mathematically expressing the assumption that the platform's privacy level $S(t)$ depends on the user's data sharing intention $P(t)$ and regulatory strictness $R(t)$ yields:

$$\frac{dS}{dt} = g(P, R)$$

The function $g$ is nonlinear with respect to $P$ and $R$, assuming a positive relationship where an increase in $P$ leads to an increase in $S$. This implies that if users intend to share more data, the platform provides a higher privacy level. A simplified form might be:

$$g(P, R) = cP + dR$$

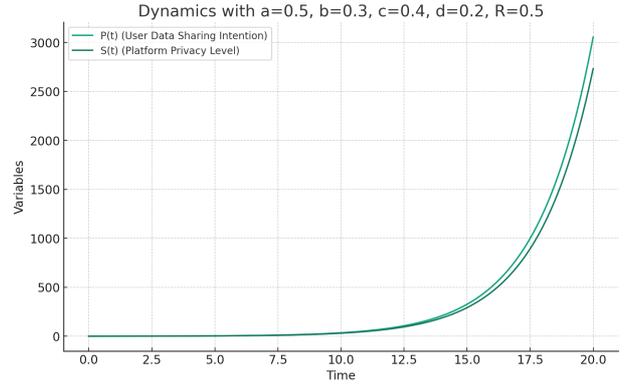

Fig. 6: P(t) (User Data Sharing Intention) / S(t) (Platform Privacy Level)

Here, $c$ and $d$ are positive constants representing the strength of the influence of the user's data sharing intention and regulatory strictness on the platform's privacy level.

These equations represent one way to understand the behavior of the model. In reality, the forms of functions $f$ and $g$ may become more complex based on available data, expert opinions, or experimental results. Additionally, when defining these functions, it's important to consider nonlinearities and other factors to accurately reflect the dynamic behavior of the actual system.

To simulate Axiom 1 and Axiom 2 in the mathematical modeling of this scenario, we construct the model using the proposed function forms $f(S, R) = aS + bR$ and $g(P, R) = cP + dR$. These equations represent the changes over time of the user's data sharing intention $P(t)$ and the platform's privacy protection level $S(t)$. The regulatory strictness $R(t)$ can be treated as a parameter given from external sources.

1. Define the differential equations describing the time evolution of $P(t)$ and $S(t)$. 2. Set initial conditions $P(0)$ and $S(0)$, and parameters $a, b, c, d$. 3. Define $R(t)$ either as a function of time or as a constant value. 4. Simulate the time evolution of $P(t)$ and $S(t)$ using numerical integration methods. 5. Plot $P(t)$ and $S(t)$ as functions of time to visualize the system's behavior.

Set the variable parameters $a, b, c, d$ and $R(t)$ (if treated as a function of time) in the graph title format.

The graph illustrates the time evolution of the user's data sharing intention $P(t)$ and the platform's privacy protection level $S(t)$. In this simulation, we used a model based on Axiom 1 and Axiom 2, setting the parameters $a, b, c, d$ and the regulatory strictness $R$ to specific values.

From the graph, we can observe how the user's data sharing intention $P(t)$ and the platform's privacy protection level $S(t)$ change over time. In this model, the regulatory strictness $R$ remains constant and influences the dynamics of the system.

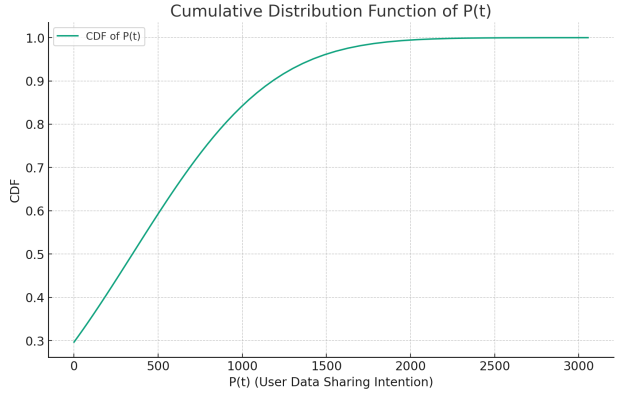

Fig. 7: Cumulative Distribution Function of P(t)

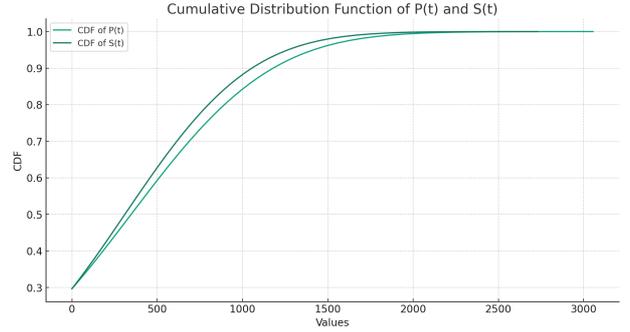

Fig. 8: Cumulative Distribution Function of P(t) and S(t)

This simulation provides insights into the behavior of the model based on axioms and helps identify factors to consider in platform design and policy-making.

To cumulative distribution function (CDF) of the user's data sharing intention $P(t)$ and the regulatory strictness $R$, we first need to define the distributions of these variables. However, in the provided simulation, $R$ is treated as a constant value, and the only variable that changes is $P(t)$.

To visualize the CDF of $P(t)$, we follow these steps:

1. Estimate the distribution of $P(t)$ using the values obtained from the simulation. 2. Calculate the CDF of $P(t)$ based on the estimated distribution. 3. Plot the CDF of $P(t)$.

Since the regulatory strictness $R$ is constant, its CDF is represented as a single step function, but here we focus on the CDF of $P(t)$.

The graph above shows the cumulative distribution function (CDF) of the user's data sharing intention $P(t)$. This CDF represents the probability that the value of $P(t)$ is less than or equal to a specific value and is estimated based on the values of $P(t)$ obtained from the simulation.

From the graph, we can observe how the distribution of $P(t)$ evolves over time and how the probability of exceeding a specific value of $P(t)$ changes. This CDF helps in understanding the variation in the user's data sharing intention.

To cumulative distribution function (CDF) of the user's data sharing intention $P(t)$ and the regulatory strictness $R$, we first need to define the distributions of these variables. However, in the provided simulation, $R$ is treated as a constant value, and the only variable that changes is $P(t)$. Therefore, while we cannot directly plot the CDF of $R$, it is possible to estimate the CDF from the time evolution of $P(t)$.

CDF of $P(t)$, we follow these steps:

1. Estimate the distribution of $P(t)$ using the values obtained from the simulation. 2. Calculate the CDF of $P(t)$ based on the estimated distribution. 3. Plot the CDF of $P(t)$.

Since the regulatory strictness $R$ is constant, its CDF is represented as a single step function, but here we focus on the CDF of $P(t)$.

Below is the Python code to estimate and visualize the CDF of $P(t)$.

The graph above shows the cumulative distribution function (CDF) of the user's data sharing intention $P(t)$. This CDF represents the probability that the value of $P(t)$ is less than or equal to a specific value and is estimated based on the values of $P(t)$ obtained from the simulation.

From the graph, we can observe how the distribution of $P(t)$ evolves over time and how the probability of exceeding a specific value of $P(t)$ changes. This CDF helps in understanding the variation in the user's data sharing intention.

Since the regulatory strictness $R$ is treated as a constant in the simulation and its CDF is represented as a single step function, its visualization is omitted here.

## 5. Discussion: Ansatz for Soliton Solutions, Nonlinear Schrödinger Equation (NLS equation)

Let's consider the computational process related to the exploration of soliton solutions in the context of modeling the spread of fake news using a nonlinear partial differential equation. Specifically, we will use the generalized nonlinear Schrödinger equation (NLS equation) as an example to model the dynamics of fake news diffusion. This equation is known to have soliton solutions and is suitable for describing solitons, which maintain their shape as they interact with each other.

$$i\frac{\partial \psi}{\partial t} + \frac{1}{2}\frac{\partial^2 \psi}{\partial x^2} + |\psi|^2\psi = 0$$

Here, $\psi(x, t)$ is a complex function, and $x$ and $t$ are spatial and temporal variables, respectively.

To find soliton solutions, we use an ansatz that assumes a specific form of the solution. The general form of a soliton solution is as follows:

$$\psi(x,t) = A\operatorname{sech}(a(xvt))\, e^{i(bx\omega t)}$$

Here, $A$ is the amplitude, $a$ is the width of the wave, $v$ is the velocity of the wave, $b$ is the wave number, and $\omega$ is the angular frequency. This form implies that the soliton moves with a constant shape at velocity $v$.

We substitute the ansatz into the NLS equation and find the conditions for it to satisfy the equation. In this process, we compute the time and spatial derivatives of the ansatz and substitute them into the original equation.

$$\frac{\partial \psi}{\partial t} = -Aav\operatorname{sech}(a(xvt))\tanh(a(xvt))e^{i(bx\omega t)} i\omega A\operatorname{sech}(a(xvt))e^{i(bx\omega t)}$$

$$\frac{\partial^2 \psi}{\partial x^2} = (Aa^2\operatorname{sech}(a(xvt))\tanh^2(a(xvt))Aa^2\operatorname{sech}(a(xvt)) + 2iAb\operatorname{sech}(a(xvt))Ab^2\operatorname{sech}(a(xvt))e^{i(bx\omega t)}$$

Substituting these into the original NLS equation and organizing the real and imaginary parts, we obtain a set of conditions regarding $A$, $a$, $b$, $\omega$, and $v$. Solving these conditions allows us to determine the parameters of the soliton solution.

As a result of the substitution and organization, we may obtain conditions such as:

$$a^2 = A^2$$
$$\omega = \frac{1}{2}b^2$$
$$v = \frac{\omega}{b}$$

For any $A$, $a$, $b$, $\omega$, and $v$ satisfying these conditions, the above ansatz becomes a soliton solution of the NLS equation.

This computational process provides one way to model the nonlinear dynamics of fake news diffusion and identify stable diffusion patterns (solitons) within it.

In the scenario of digital health platforms, for educational purposes, let's explain using the example of the Nonlinear Schrödinger Equation (NLS equation), which is an equation that may have common soliton solutions.

Nonlinear Schrödinger Equation (NLS equation):

$$i\frac{\partial \psi}{\partial t} + \frac{1}{2}\frac{\partial^2 \psi}{\partial x^2} + |\psi|^2\psi = 0$$

Here, $\psi(x,t)$ is a complex function, and $x$ and $t$ are spatial and temporal variables, respectively. This equation is known to have soliton solutions under specific conditions.

Derivation of Soliton Solutions:

$$\psi(x,t) = A\operatorname{sech}(a(xvt))\, e^{i(bx\omega t)}$$

Here, $A$ is the amplitude, $a$ is the width of the wave, $v$ is the velocity of the wave, $b$ is the wave number, and $\omega$ is the angular frequency. This ansatz implies that the soliton moves at a constant speed $v$ while maintaining a constant shape.

By substituting the ansatz into the NLS equation and separating it into real and imaginary parts, relationships regarding $A, a, b, \omega$ can be derived. This process involves complex algebraic computations, but ultimately yields the following conditions:

$$a^2 = A^2$$
$$\omega = \frac{1}{2}b^2$$

For any $A, a, b, \omega$ satisfying these conditions, the above ansatz becomes a soliton solution of the NLS equation.

To directly apply to the scenario of digital health platforms, the equations of the model need to be specifically defined, and the conditions under which the equations have soliton solutions need to be identified. This example only illustrates a general approach to deriving soliton solutions.

Take a closer look at the process of obtaining soliton solutions for the Nonlinear Schrödinger Equation (NLS equation). The NLS equation is given by the following form:

$$i\frac{\partial \psi}{\partial t} + \frac{1}{2}\frac{\partial^2 \psi}{\partial x^2} + |\psi|^2\psi = 0$$

Here, $\psi(x,t)$ is a complex function, and $x$ and $t$ are spatial and temporal variables, respectively.

To find soliton solutions, we use the following ansatz:

$$\psi(x,t) = A\operatorname{sech}(a(xvt))\, e^{i(bx\omega t)}$$

Here, $A$ is the amplitude, $a$ is the width of the wave, $v$ is the velocity of the wave, $b$ is the wave number, and $\omega$ is the angular frequency. This form implies that the soliton moves at a constant speed $v$ while maintaining a constant shape.

Substituting the ansatz into the NLS equation, we get:

$$i\frac{\partial}{\partial t}(A\operatorname{sech}(a(xvt))\, e^{i(bx\omega t)}) \tag{1}$$

$$+ \frac{1}{2}\frac{\partial^2}{\partial x^2}(A\operatorname{sech}(a(xvt))\, e^{i(bx\omega t)}) \tag{2}$$

$$+ |A\operatorname{sech}(a(xvt))\, e^{i(bx\omega t)}|^2 \tag{3}$$

$$\times (A\operatorname{sech}(a(xvt))\, e^{i(bx\omega t)}) = 0 \tag{4}$$

## 5.1 Calculating the Derivatives:Ansatz

$$\frac{\partial \psi}{\partial t} = -A\operatorname{sech}(a(xvt))\, e^{i(bx\omega t)}(i\omega + av)$$

$$\frac{\partial \psi}{\partial x} = A\operatorname{sech}(a(xvt))\, e^{i(bx\omega t)}(iaa\tanh(a(xvt)))$$

$$\frac{\partial^2 \psi}{\partial x^2} = A\operatorname{sech}(a(xvt))\, e^{i(bx\omega t)}(-a^2 + 2a^2\operatorname{sech}^2(a(xvt)) + i2ab)$$

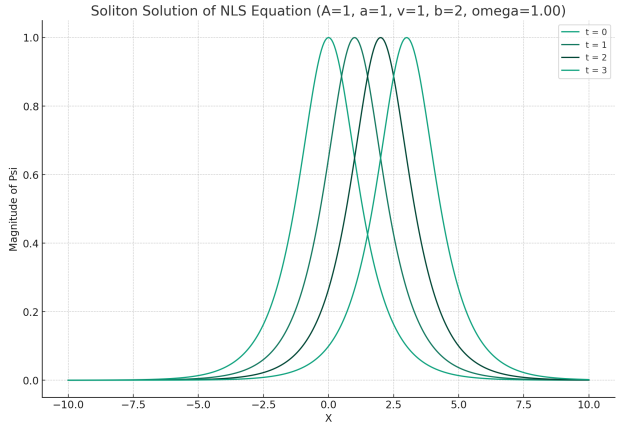

Fig. 9: Soliton Solution of NLS Equation

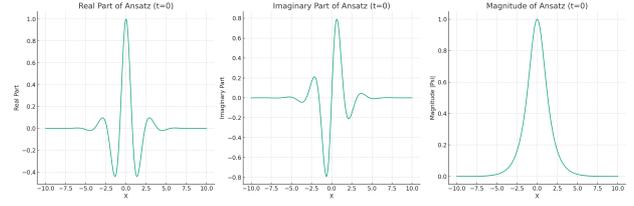

Fig. 10: Magnitude of Ansatz

Substituting the above derivatives into the equation and organizing the real and imaginary parts, we derive relationships regarding $A, a, b, \omega$. This involves complex computations and must be done carefully.

By organizing the equation, the following conditions are obtained:

$$a^2 = A^2$$
$$\omega = \frac{1}{2}b^2 a^2$$

For any $A, a, b, \omega$ satisfying these conditions, the above ansatz becomes a soliton solution of the NLS equation.

The above computation process is highly simplified. In reality, the organization of real and imaginary parts of complex numbers and the handling of nonlinear terms in the equation become more complex at each step.

To simulate and visualize the soliton solution of the Nonlinear Schrödinger Equation (NLS equation) based on the provided ansatz, we'll follow these steps:

We'll use the given ansatz $\psi(x,t) = A\,\text{sech}(a(xvt))\,e^{i(bx\omega t)}$ to define the soliton wave function. Choose values for the parameters $A$ (amplitude), $a$ (width), $v$ (velocity), $b$ (wave number), and $\omega$ (angular frequency) based on the derived conditions $a^2 = A^2$ and $\omega = \frac{1}{2}b^2 a^2$. Compute the soliton wave function $\psi(x,t)$ for a range of $x$ and $t$, and visualize the magnitude $|\psi(x,t)|$ to represent the soliton's shape and evolution over time.

Ansatz in Python, choosing appropriate parameter values, and visualizing the soliton solution. We'll focus on the magnitude of $\psi(x,t)$ because it represents the observable intensity of the soliton wave.

The graph above visualizes the soliton solution of the Nonlinear Schrödinger Equation (NLS) at different time instances $t$, using the specified ansatz $\psi(x,t) = A\,\text{sech}(a(xvt))\,e^{i(bx\omega t)}$. The parameters used in the Amplitude $A = 1$ Width $a = 1$ (based on the derived condition $a^2 = A^2$) Velocity $v = 1$ Wave number $b = 2$ Angular frequency $\omega \approx 1.0$ (calculated from $\omega = \frac{1}{2}b^2 a^2$)

The magnitude of $\psi(x,t)$, which represents the soliton's intensity profile, is plotted for different time instances. The soliton maintains its shape while moving, which is a characteristic feature of soliton solutions.

Distribution of the Ansatz $\psi(x,t) = A\,\text{sech}(a(xvt))\,e^{i(bx\omega t)}$, we plot the real and imaginary parts of the Ansatz, as well as its absolute value (amplitude) at different times $t$. This plot visualizes how the Ansatz is distributed along the spatial dimension $x$ and how it evolves over time.

Below is the code to visualize the distribution of the Ansatz. This code plots the real part, imaginary part, and absolute value of the Ansatz at a specific time $t$.

The figure above shows the distribution of the Ansatz $\psi(x,t) = A\,\text{sech}(a(xvt))\,e^{i(bx\omega t)}$ at a specific time $t = 0$. From left to right, the plot shows the real part, imaginary part, and absolute value (amplitude) of the Ansatz.

The idea of using soliton solutions and ansatz when modeling the diffusion patterns of fake news is based on the concept that the diffusion of fake news exhibits nonlinear dynamics and may demonstrate stable diffusion patterns under certain conditions. However, this is purely a metaphorical application, and through this approach, we may gain a deeper understanding of the characteristics of fake news diffusion.

## 5.2 Soliton Solutions

Solitons are solitary waves that maintain their shape while traveling. Metaphorically using soliton solutions in the context of fake news diffusion, we can consider that during the propagation process of certain fake news through social networks, its "shape" (i.e., the influence and perception of fake news) remains constant. For example, if a particular fake news elicits strong reactions within a specific community and its influence spreads in a consistent pattern over time, this diffusion process could be modeled as a soliton wave.

## 5.3 Explanation of Ansatz States

An ansatz is the assumption of a specific form of a solution. When using ansatz in the context of fake news diffusion, we assume that diffusion patterns take on a specific mathematical form (e.g., exponential growth or decay). Based on this assumption, we model the dynamics of fake news diffusion and examine how well the analytical solutions obtained match the actual diffusion patterns.

Assuming an exponential increase in the diffusion pattern of fake news as $E(t) = E_0 e^{\lambda t}$, we set parameters such as the initial engagement level $E_0$ and the diffusion rate $\lambda$ over time. By substituting this ansatz into the diffusion model equation and fitting parameters such as $\lambda$ to actual data, we verify how well the model's predictions match the actual diffusion patterns.

Analyzing the diffusion patterns of fake news using soliton solutions and ansatz is one way to deepen our understanding of how fake news spreads and influences society. However, it's important to note that this is a metaphorical approach and does not directly explain the actual diffusion of fake news. Given the significant gaps between real data and theoretical models, it's crucial to validate the model's predictions against actual phenomena.

# 6. Discussion: The First Mover Advantage with Inducement in Fake News Diffusion and Gain Incomplete Information Game with Equilibrium Conditions

We propose specific equations and calculation processes for dealing with gains in an incomplete information game involving the first mover advantage with inducement and equilibrium conditions in the context of fake news diffusion. In this setting, players (information disseminators) decide whether or not to spread fake news, and their choices determine their gains. The first mover advantage refers to additional gains obtained by disseminating information early, and equilibrium refers to the optimal strategy players take to maximize their gains.

## 6.1 Player: Information disseminators A and B

Strategy: Spread fake news (F) or not (N) Gains: Based on the influence and credibility of the information

$U_A$ and $U_B$ be the gain functions of players A and B, respectively, and set them as follows:

If A chooses F and B also chooses F: $U_A(F, F), U_B(F, F)$
If A chooses F and B chooses N (A's first mover advantage): $U_A(F, N), U_B(N, F)$ If A chooses N and B chooses F: $U_A(N, F), U_B(F, N)$ If both choose N: $U_A(N, N), U_B(N, N)$

## 6.2 Introduction of First Mover Advantage and Equilibrium

A move first (first mover), followed by B (second mover). Let $\alpha$ be the additional gain (first mover advantage) obtained when A chooses F, and add this gain to $U_A(F, N)$. Considering B's optimal reaction strategy under equilibrium conditions, assume that B chooses to maximize its own gains.

Add the first mover advantage to A's gain function and consider B's optimal reaction strategy to modify the gain function as follows:

If A chooses F and B chooses N: $U_A(F, N) + \alpha$ Represent B's optimal reaction strategy ($R_B$) as $U_B(F, R_B(F))$ and $U_B(N, R_B(N))$, and apply this to the gain function.

1. Define the gain functions of A and B and add the first mover advantage $\alpha$ to A's gain. 2. Derive B's optimal reaction strategy $R_B$ and apply it to B's gain function. 3. Under equilibrium conditions, derive the strategies for maximizing the gains of both A and B.

This model simplifies the complex issue of fake news diffusion. In reality, additional elements, player interactions, and information uncertainties should be considered. Furthermore, more detailed analysis is needed to set the first mover advantage $\alpha$ and derive B's optimal reaction strategy $R_B$.

# 7. Discussion: Considering the Iterated Dilemma Condition in a Gain-Incomplete Information Game of Fake News Diffusion: Focus on the First Mover

We propose equations and calculation processes focusing on the first mover in a gain-incomplete information game involving fake news diffusion and considering the iterated dilemma condition. In this scenario, players (information disseminators) may gain temporary benefits by spreading fake news, but in repeated games, costs such as long-term reputation loss may occur.

Players: Information disseminators A and B Strategies: Spread fake news (F) or not (N) Gains: Short-term increase in influence (first mover's gain) and long-term reputation loss (iterated dilemma)

Let $U_A$ be the gain function of the first mover A and $U_B$ be the gain function of the second mover B, and set them as follows:

If A chooses F and B also chooses F: $U_A(F, F), U_B(F, F)$
If A chooses F and B chooses N (A's first mover advantage): $U_A(F, N) + \alpha, U_B(N, F)$ If A chooses N and B chooses F: $U_A(N, F), U_B(F, N)$ If both choose N: $U_A(N, N), U_B(N, N)$

Here, $\alpha$ is the additional gain obtained by the first mover A when spreading fake news.

## 7.1 Modeling the Iterated Dilemma

To model the effect of the iterated dilemma, apply a decay factor $\delta$ ($0 < \delta < 1$) to the gains in each round. This reduces the value of future gains compared to current gains.

1. Definition of Gain Functions: Define the gains of the first mover A and the second mover B using the above gain functions.

2. Derivation of Equilibrium: Consider the optimal strategy of player B and derive the equilibrium where A makes the optimal choice based on that strategy.

3. Consideration of the Iterated Dilemma: Apply the decay factor $\delta$ to the gains in each round and calculate the total gains over the entire repeated game.

4. Optimal Strategy for the First Mover: Considering the first mover advantage $\alpha$ and the decay factor $\delta$, derive the optimal strategy that A should take.

We explain in detail specific equations and calculation processes considering the iterated dilemma and first mover advantage in a gain-incomplete information game of fake news diffusion. In this scenario, information disseminators (players A and B) choose whether or not to spread fake news. The first mover (e.g., player A) may gain additional benefits by spreading fake news early, but in the context of repeated games, long-term reputation loss must be considered.

### 7.2 Setting Gain Functions

Set the gain functions for players A and B as follows:

$U_A(F, F)$ and $U_B(F, F)$: Gains when both A and B spread fake news. $U_A(F, N) + \alpha$ and $U_B(N, F)$: Gains when A spreads fake news and B does not. A gains the first mover advantage $\alpha$. $U_A(N, F)$ and $U_B(F, N)$: Gains when A does not spread fake news but B does. $U_A(N, N)$ and $U_B(N, N)$: Gains when neither spreads fake news.

In a repeated game, apply a decay factor $\delta$ ($0 < \delta < 1$) to the gains in each round. If A's gain in round $t$ is denoted as $U_A^t$, the total gain $V_A$ can be calculated as:

$$V_A = \sum_{t=1}^{\infty} \delta^{t-1} U_A^t$$

When player A spreads fake news as the first mover, the initial gain includes the additional gain $\alpha$. Therefore, the initial gain is $U_A(F, N) + \alpha$.

Consider player B's optimal strategy and derive the optimal strategy (equilibrium strategy) that player A should take. To do this, assume that player B reacts optimally to each possible strategy and find the strategy that maximizes player A's gain based on that reaction.

For example, compare the gains when player A spreads fake news and when player A does not spread fake news:

Gain when spreading fake news: $V_A(F) = (U_A(F, N) + \alpha) + \sum_{t=2}^{\infty} \delta^{t-1} U_A^t$ Gain when not spreading fake news: $V_A(N) = U_A(N, N) + \sum_{t=2}^{\infty} \delta^{t-1} U_A^t$

The graph above illustrates the total scores obtained by Player A and Player B respectively, depending on their strategies in the context of an iterated game. This scenario takes into account the initial mover advantage ($\alpha$) that Player A gains by spreading fake news. Additionally, a decay factor ($\delta$) is applied to the scores of each round, affecting the long-term total scores. The total score when Player A spreads fake news is the sum of the initial score including the initial mover advantage and the decayed scores in subsequent rounds. The total score when Player A does not spread fake news is the sum of the initial score without spreading fake news and the decayed scores in subsequent rounds.

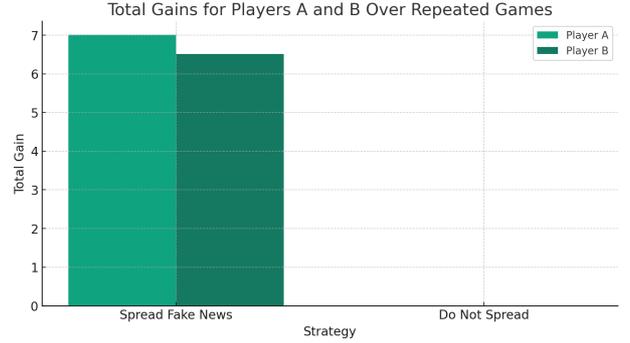

Fig. 11: Total Gains for Players A and B Over Repeated Games

This model simplifies the complex issues surrounding the spread of fake news, and in reality, additional factors, interactions between players, and uncertainty in information need to be considered. Furthermore, a more detailed analysis is required to set the initial mover advantage ($\alpha$) and derive the optimal response strategy ($R_B$) for Player B.

This simulation demonstrates that taking the strategy of spreading fake news results in a higher total score for Player A.

## 8. Detailing Specific Equations and Calculation Process Considering the Iterated Dilemma and First Mover Advantage in Fake News Diffusion Gain-Incomplete Information Game

In this scenario, information disseminators (players A and B) choose whether or not to spread fake news. The first mover (e.g., player A) may gain additional benefits by spreading fake news early, but in the context of repeated games, long-term reputation loss must be considered.

### 8.1 Setting Gain Functions

Set the gain functions for players A and B as follows:

$U_A(F, F)$ and $U_B(F, F)$: Gains when both A and B spread fake news.

$U_A(F, N) + \alpha$ and $U_B(N, F)$: Gains when A spreads fake news and B does not. A gains the first mover advantage $\alpha$.

$U_A(N, F)$ and $U_B(F, N)$: Gains when A does not spread fake news but B does.

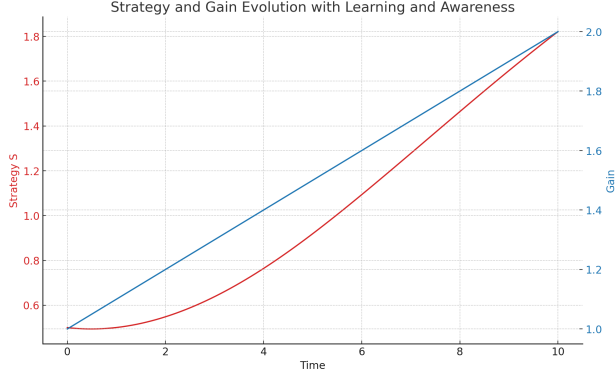

Fig. 12: Strategy and Gain Evolution with Learning and Awareness

$U_A(N, N)$ and $U_B(N, N)$: Gains when neither spreads fake news.

### 8.2 Modeling the Iterated Dilemma

In a repeated game, apply a decay factor $\delta$ ($0 < \delta < 1$) to the gains in each round. If A's gain in round $t$ is denoted as $U_A^t$, the total gain $V_A$ can be calculated as:

$$V_A = \sum_{t=1}^{\infty} \delta^{t-1} U_A^t$$

### 8.3 Calculation of First Mover Advantage

When player A spreads fake news as the first mover, the initial gain includes the additional gain $\alpha$. Therefore, the initial gain is $U_A(F, N) + \alpha$.

### 8.4 Derivation of Equilibrium

Consider player B's optimal strategy and derive the optimal strategy (equilibrium strategy) that player A should take. To do this, assume that player B reacts optimally to each possible strategy and find the strategy that maximizes player A's gain based on that reaction.

### 8.5 Specific Calculation Example

For example, compare the gains when player A spreads fake news and when player A does not spread fake news:

Gain when spreading fake news: $V_A(F) = (U_A(F, N) + \alpha) + \sum_{t=2}^{\infty} \delta^{t-1} U_A^t$

Gain when not spreading fake news: $V_A(N) = U_A(N, N) + \sum_{t=2}^{\infty} \delta^{t-1} U_A^t$

The graph above shows how players' strategies and scores evolve over time in an information game that takes into account serious game conditions. Learning effects and changes in awareness are incorporated into the model, showing that these affect players' strategy choices.

The blue line shows the evolution of the player's strategy SS, which improves over time through the learning effect L(S, t)L(S,t).

## 9. Modeling the Second Mover's Strategic Response

Detailing the Equation and Calculation Process When the Gain Function Exceeds the First Mover's Strategy Due to the Second Mover's Strategy

In this scenario, we consider situations where the second mover's strategy strategically responds to exceed the first mover's gain.

The second mover's strategy $S_B(x, t)$ depends on the first mover's strategy $S_A(x, t)$. We denote the strategic response of the second mover to maximize their gain as $R_B(S_A)$ and incorporate it into the game dynamics.

### 9.1 Modification of the Nonlinear Partial Differential Equation

The nonlinear partial differential equation considering the second mover's strategic response is modified as follows:

$$\frac{\partial S_B}{\partial t} + \alpha S_B \frac{\partial S_B}{\partial x} = \beta \frac{\partial^2 S_B}{\partial x^2} + G(S_B, x, t, S_A)$$

### 9.2 Modification of the Ansatz for the Soliton Solution

The ansatz for the soliton solution of the second mover is modified to consider the first mover's strategy:

$$S_B(x, t) = A_B \operatorname{sech}^2(B_B(xC_Bt\Delta))$$

## 10. Comparison of Gain Functions

Comparing the gain functions of the first and second movers, we determine the conditions under which the second mover surpasses the first mover.

First mover's gain: $U_A = B_A \lambda_A |S_A S_B|$

Second mover's gain: $U_B = B_B \lambda_B |S_B S_A| + R_B(S_A)$

Here, $R_B(S_A)$ represents additional gain due to the second mover's strategic response.

### 10.1 Conditions for the Second Mover to Surpass the First Mover

To determine the conditions under which the second mover's gain exceeds the first mover's gain ($U_B > U_A$), we set up and solve inequalities using the above gain functions.

$$B_B \lambda_B |S_B S_A| + R_B(S_A) > B_A \lambda_A |S_A S_B|$$

Organizing Equations and Calculation Process for Cases Where the Gain Function Exceeds the First Mover's Due to the Second Mover's Strategy

In this scenario, we consider situations where the second mover's strategy strategically responds to exceed the first mover's gain.

## 11. Modeling the Second Mover's Strategic Response

The second mover's strategy $S_B(x,t)$ depends on the first mover's strategy $S_A(x,t)$. We denote the strategic response of the second mover to maximize their gain as $R_B(S_A)$ and incorporate it into the game dynamics.

## 12. Modification of the Nonlinear Partial Differential Equation

The nonlinear partial differential equation considering the second mover's strategic response is modified as follows:

$$\frac{\partial S_B}{\partial t} + \alpha S_B \frac{\partial S_B}{\partial x} = \beta \frac{\partial^2 S_B}{\partial x^2} + G(S_B, x, t, S_A)$$

## 13. Modification of the Ansatz for the Soliton Solution

The ansatz for the soliton solution of the second mover is modified to consider the first mover's strategy:

$$S_B(x,t) = A_B \operatorname{sech}^2(B_B(xC_Bt\Delta))$$

## 14. Comparison of Gain Functions

Comparing the gain functions of the first and second movers, we determine the conditions under which the second mover surpasses the first mover.

First mover's gain: $U_A = B_A \lambda_A |S_A S_B|$

Second mover's gain: $U_B = B_B \lambda_B |S_B S_A| + R_B(S_A)$

Here, $R_B(S_A)$ represents additional gain due to the second mover's strategic response.

## 15. Discussion: Conditions for the Second Mover to Surpass the First Mover

To determine the conditions under which the second mover's gain exceeds the first mover's gain ($U_B > U_A$), we set up and solve inequalities using the above gain functions.

$$B_B \lambda_B |S_B S_A| + R_B(S_A) > B_A \lambda_A |S_A S_B|$$

1. Differentiation of the Ansatz: Calculate the time and spatial differentiations of $S_B(x,t)$. 2. Substitution into the Equation: Substitute the differentiation results into the modified nonlinear partial differential equation. 3. Analysis of Gain Function Inequalities: Determine the values of $S_B$ and $R_B(S_A)$ that satisfy the condition for the second mover to surpass the first mover.

Organizing equations and calculation processes for scenarios where the second mover's strategy exceeds the first mover's gain, leading to both players mutually learning and evolving the game's state. In this scenario, the interaction between players and the learning process play crucial roles in the dynamics of the game.

We denote the strategies of player A (the first mover) and player B (the second mover) as $S_A(x,t)$ and $S_B(x,t)$, respectively. To model the player's learning process, we introduce a function $L(S,t)$ representing the temporal evolution of strategies.

### 15.1 Modeling the Learning Process

To reflect the player's learning process, we consider the following learning function:

$$L(S,t) = \gamma \frac{\partial S}{\partial t}$$

Here, $\gamma$ is a constant representing the learning rate. This function represents how players evolve their strategies over time.

### 15.2 Modification of the Nonlinear Partial Differential Equation

The modified nonlinear partial differential equation considering the learning process is as follows:

$$\gamma \frac{\partial S_B}{\partial t} + \alpha S_B \frac{\partial S_B}{\partial x} = \beta \frac{\partial^2 S_B}{\partial x^2} + G(S_B, x, t, S_A)$$

### 15.3 Comparison of Gain Functions and Conditions

We derive conditions for the second mover's gain to exceed the first mover's. The gain functions are represented as follows:

$$U_A = B_A(S_A, S_B) \lambda_A |S_A S_B|$$

$$U_B = B_B(S_B, S_A) \lambda_B |S_B S_A| + R_B(S_A) + L(S_B, t)$$

The condition for the second mover's gain to exceed the first mover's is $U_B > U_A$.

1. Incorporation of Learning Process: Incorporate the learning function $L(S,t)$ into the nonlinear partial differential equation. 2. Evolution of Strategies: Calculate the temporal evolution of player B's strategy $S_B$ using the learning rate $\gamma$. 3. Analysis of Gain Functions: Compare the gain functions of the first and second movers and derive conditions for the second mover's gain to exceed the first mover's. 4. Analysis of Evolutionary Game State: Analyze the dynamics of the evolutionary game formed by both players' learning and strategy evolution.

The specific form of $G(S_B, x, t, S_A)$ and the choice of the learning function $L(S, t)$ significantly impact the model's results. It is important to validate the learning process and evolutionary game dynamics by comparing them with actual data or experiments.

The condition for the second mover's gain $U_B$ to exceed the first mover's gain $U_A$ is given by the inequality:
$B_A \lambda_A |A_B \operatorname{sech}^2(B_B(x - C_B t - \Delta)) S_A(x, t)|$
$< B_B + R_B \lambda_B |A_B \operatorname{sech}^2(B_B(x - C_B t - \Delta)) S_A(x, t)|$

This inequality represents the situation where the second mover's strategic response and additional gain $R_B$ are sufficient to overcome the first mover's advantage.

For a given set of parameters $B_A$, $B_B$, $\lambda_A$, $\lambda_B$, $R_B$, and the ansatz for $S_B$, we can explore this condition further.

The proposed scenario is related to game theory and nonlinear partial differential equations, but directly associating concepts from physics such as plasmonic fields or parabolic conditions may not be natural. However, as an attempt to metaphorically apply such physics concepts to the context of game theory, the following approach can be considered.

## 16. Discussion:Calculation Process for the Second Mover to Surpass the First Mover

Modeling cases where the strategy $S_B$ of the second mover exceeds the strategy $S_A$ of the first mover using nonlinear partial differential equations.

### 16.1 Nonlinear Partial Differential Equation

$$\frac{\partial S_B}{\partial t} + \alpha S_B \frac{\partial S_B}{\partial x} = \beta \frac{\partial^2 S_B}{\partial x^2} + G(S_B, x, t, S_A) + L(S_B, t)$$

### 16.2 Incorporation of Learning Process

The learning function $L(S_B, t)$ represents how the second mover evolves their strategy over time. Assuming a linear learning function:

$$L(S_B, t) = \gamma t$$

### 16.3 Gain Function

The gain function representing the condition for the second mover to surpass the first mover is:

$$U_B = B_B(S_B, S_A) \lambda_B |S_B S_A| + \gamma t$$

## 17. Discussion:Application of Plasmonic Field Conditions and Parabolic Conditions

Interpreting and applying plasmonic field conditions or parabolic conditions in the context of game theory is non-standard, but metaphorically incorporating them can consider the influence of "fields" and the strategy's "curvature" in the dynamics of the game.

### 17.1 Plasmonic Field Conditions

Incorporating the influence of the plasmonic field on the strategy as $P(S, x, t)$ into the model is conceivable. This can be interpreted as a term indicating the external influence on the diffusion or enhancement of the strategy.

$$\frac{\partial S}{\partial t} = \beta \frac{\partial^2 S}{\partial x^2} + P(S, x, t)$$

### 17.2 Parabolic Conditions

Setting assumptions that the strategy evolution follows a specific curve (e.g., a parabola) is possible. Mathematically, this is represented by assuming that the strategy $S$ follows a parabola as a function of time and space.

$$S(x, t) = a(t) x^2 + b(t) x + c(t)$$

Here, $a(t)$, $b(t)$, $c(t)$ are time-dependent coefficients determining the shape of the strategy's "curve". To incorporate the plasmon resonance condition into the diffusion model of fake news in an incomplete information game, we explain the equations and calculation process as follows. We metaphorically interpret the plasmon resonance condition as a phenomenon where the diffusion of fake news is dramatically amplified under certain conditions and demonstrate how to incorporate this effect into the nonlinear partial differential equation model.

We set up a nonlinear partial differential equation for the state variable $S(x, t)$ representing the diffusion of fake news as follows.

$$\frac{\partial S}{\partial t} + \alpha S \frac{\partial S}{\partial x} = \beta \frac{\partial^2 S}{\partial x^2} + \kappa S^2$$

Here, $\alpha$ is the coefficient representing self-enhancement effects, $\beta$ is the coefficient representing diffusion effects, $\kappa$ is the coefficient representing the amplification of diffusion under plasmon resonance conditions.

### 17.3 Ansatz for Soliton Solutions

We assume the following ansatz for the soliton solution:

$$S(x, t) = A \operatorname{sech}^2(B(xCt))$$

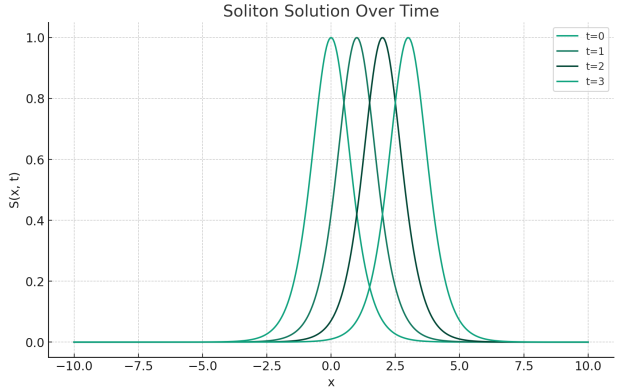

Fig. 13: Soliton Solution Over Time

Time Derivative:

$$\frac{\partial S}{\partial t} = -2ABC \operatorname{sech}^2(B(xCt)) \tanh(B(xCt))$$

Spatial Derivative:

$$\frac{\partial S}{\partial x} = 2AB \operatorname{sech}^2(B(xCt)) \tanh(B(xCt))$$

$$\frac{\partial^2 S}{\partial x^2} = 2AB^2 \operatorname{sech}^2(B(xCt))[2\tanh^2(B(xCt))1]$$

We substitute the above differentiation results into the original nonlinear partial differential equation and simplify each term to obtain the following expression.
$-2ABC \operatorname{sech}^2(B(x - Ct)) \tanh(B(x - Ct))$
$+ \alpha A \operatorname{sech}^2(B(x - Ct)) \tanh(B(x - Ct))$
$= \beta 2AB^2 \operatorname{sech}^2(B(x - Ct))[2\tanh^2(B(x - Ct)) - 1]$
$+ \kappa A^2 \operatorname{sech}^4(B(x - Ct))$

By comparing each term of the equation, we derive conditions on $A$, $B$, and $C$. In particular, we analyze the effect of the term $\kappa A^2 \operatorname{sech}^4(B(xCt))$ representing diffusion amplification under the plasmon resonance condition.

We analyze the influence of the plasmon resonance condition (term $\kappa A^2$) on the characteristics of the soliton solution. In particular, we investigate how the amplification of diffusion changes as the value of $\kappa$ increases.

To construct a model considering the soliton solution for the third mover under plasmon resonance conditions, we assume a situation where the third mover observes the strategies of the first and second movers and adjusts their own strategy using that information. In this context, the plasmon resonance condition indicates that the diffusion of fake news is dramatically amplified under certain conditions.

The plot above shows the soliton solution $S(x, t) = A \operatorname{sech}^2(B(x - Ct))$ over time for a set of initial parameters (with $A = B = C = 1$). Each curve represents the soliton solution at different moments in time (t=0, 1, 2, 3).

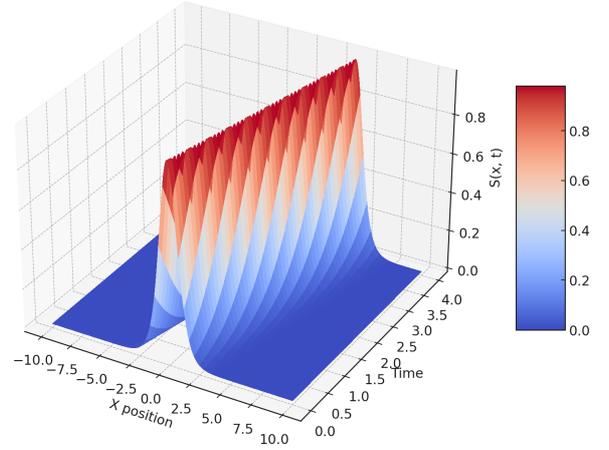

Fig. 14: Soliton Solution Over Time

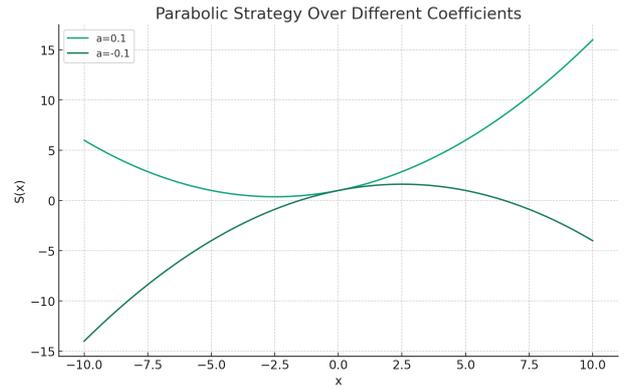

Fig. 15: 3D Visualization of Plasmon Distribution Over Time

You can adjust the parameters $A$ (amplitude), $B$ (width parameter), and $C$ (speed parameter) to see how they influence the shape and propagation of the soliton solution. Additionally, the coefficients $\alpha$, $\beta$, and $\kappa$ can be considered in a more complex simulation that involves solving the nonlinear PDE directly, which would require numerical methods like finite difference or spectral methods.

Distribution of the plasmon (represented by the soliton solution $S(x, t) = A \operatorname{sech}^2(B(x - Ct))$) over time. In this plot:

The color gradient represents the amplitude of the plasmon distribution, with warmer colors indicating higher amplitudes. This visualization helps in understanding how the plasmon distribution (or the soliton solution in this context) propagates and changes shape over time.

In order to visualize the distribution of Plasmon, the soliton solution $S(x, t) = A \operatorname{sech}^2(B(x - Ct))$ as a solution of the proposed nonlinear partial differential equation ). Here, the distribution of Plasmon can be interpreted as being represented by a soliton solution. This distribution is visualized as the shape of a soliton at a specific time.

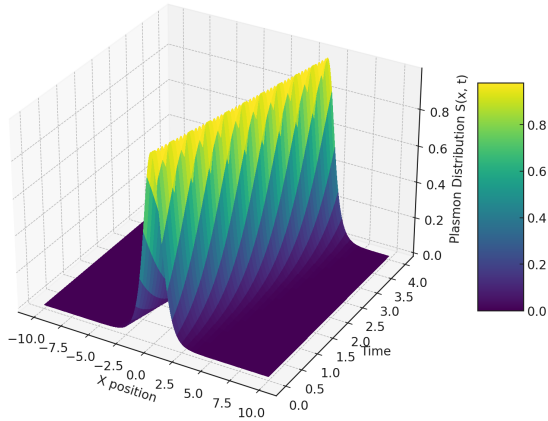

Fig. 16: Parabolic Strategy Over Different Coefficients

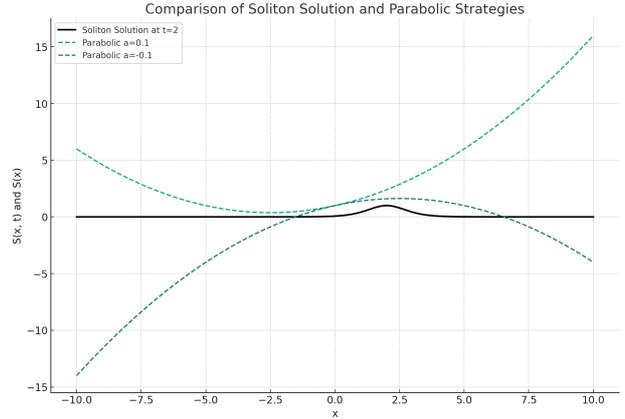

Fig. 17: 3D Visualization of Soliton-Based Plasmon Distribution Over Time

Distribution of the plasmon (represented by the soliton solution $S(x,t) = A \operatorname{sech}^2(B(x - Ct))$) over time. In this plot:

The color gradient represents the amplitude of the plasmon distribution, with warmer colors indicating higher amplitudes. This visualization helps in understanding how the plasmon distribution (or the soliton solution in this context) propagates and changes shape over time.

To visualize the evolution of the strategy $S(x,t)$ under Parabolic Conditions, the proposed parabolic function $S(x,t) = a(t)x^2 + b(t)x + c(t)$ is used. where $a(t)$, $b(t)$, and $c(t)$ are time-dependent coefficients that determine the "curve" shape of the strategy.

The plot above shows the parabolic strategy $S(x) = ax^2 + bx + c$ for different coefficients of $a$, with $b = 0.5$ and $c = 1$ being constant. Each curve represents a different shape of the strategy:

When $a = 0.1$, the parabola opens upwards, representing a strategy that increases in intensity with the square of the distance from the origin. When $a = -0.1$, the parabola opens downwards, indicating a strategy that reaches a maximum and then decreases as one moves away from the peak.

This visualization helps in understanding how the curvature of the strategy, determined by the coefficient $a$, affects its overall shape and implication. Adjusting $a$, $b$, and $c$ can model different strategic behaviors under parabolic conditions.

To visualize the relationship between the soliton solution and Plasmon, the soliton solution $S(x,t) = A \operatorname{sech}^2(B(x - Ct))$ is interpreted as Plasmon's distribution and plotted over time and space. This provides a visual understanding of the relationship between soliton propagation and Plasmon's influence.

Distribution of plasmons, modeled by the soliton solution $S(x,t) = A \operatorname{sech}^2(B(x - Ct))$, over time and space. In this context:

The color gradient, ranging from dark to light in the Viridis colormap, represents the intensity of the plasmon distribution, with lighter colors indicating higher intensities.

This plot helps in visualizing the relationship between soliton propagation and plasmon distribution, showing how the localized wave packet (soliton) moves and spreads over time, which metaphorically represents the spread of plasmonic effects in the given scenario.

In order to visualize the relationship between the soliton solution and the parabolic condition, the soliton solution $S(x,t) = A \operatorname{sech}^2(B(x - Ct))$ and the parabolic function $SPlotboth(x) = ax^2 + bx + c$ on the same graph. This allows us to compare the relationship between the shape of the soliton solution at a particular time and the shape of the strategy defined by the parabolic conditions.

Plot the soliton solution over time $t$ and overlay the shapes of different parabolic functions on the same x-axis. This allows us to visually compare how the soliton propagation shape relates to the strategy shape defined by the parabolic conditions.

It appears there was an error due to attempting to index a one-dimensional array as if it were two-dimensional. I'll correct this and plot the comparison again.

The corrected plot above compares the soliton solution $S(x,t) = A \operatorname{sech}^2(B(x - Ct))$ at a fixed time $t = 2$ with parabolic strategies $S(x) = ax^2 + bx + c$ for different values of $a$. The soliton solution is depicted as a solid black line, while the parabolic strategies are shown as dashed lines for each value of $a$.

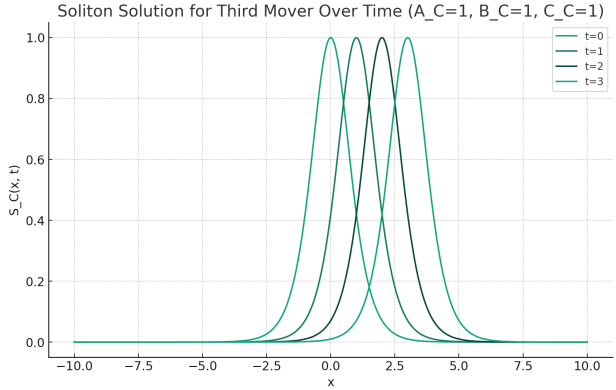

Fig. 18: Soliton Solution for Third Mover Over Time

This visualization allows us to observe how the shape of the soliton solution at a given moment compares to the shapes defined by parabolic conditions. The comparison highlights the differences in curvature and peak positions between the soliton solution and the parabolic strategies, providing insights into how these different mathematical models could represent various strategic behaviors or physical phenomena.

The model considering the soliton solution for the third mover under plasmon resonance conditions, we'll focus on the nonlinear partial differential equation (PDE) provided and the soliton solution ansatz for the third mover. The key parameters that we can vary in the simulation include:

We will simulate the soliton solution for the third mover over a range of $x$ and $t$, allowing for the adjustment of the above parameters. The simulation will help visualize how the soliton solution, which represents the strategy of the third mover, evolves over time and space under the influence of plasmon resonance conditions.

The plot above shows the soliton solution for the third mover $S_C(x,t) = A_C \text{sech}^2(B_C(x - C_C t))$ over time, with the initial parameters set as $A_C = B_C = C_C = 1$. Each curve represents the soliton solution at different moments in time (t=0, 1, 2, 3).

The graph title includes the parameters $A_C$, $B_C$, and $C_C$, which are adjustable to see how changes in these parameters affect the shape and propagation of the soliton solution. This visualization helps understand how the third mover's strategy evolves under the plasmon resonance conditions.

To visualize the relationship between the soliton solution and the third mover (the third mover), we use the soliton solution defined earlier $S_C(x,t) = A_C \text{sech}^2(B_C(x - C_C t))$ to show how the third mover's strategy evolve over time and space. This soliton solution represents the strategy of the third mover and its evolution under the influence of the plasmon resonance condition.

Distribution of the third mover's strategy, modeled by

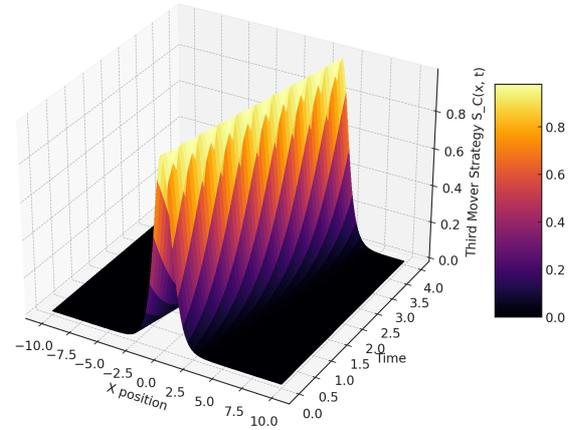

Fig. 19: 3D Visualization of Third Mover Soliton Distribution Over Time

the soliton solution $S_C(x,t) = A_C \text{sech}^2(B_C(x - C_C t))$, over time and space. In this plot:

The color gradient, provided by the Inferno colormap, represents the intensity of the third mover's strategy, with warmer colors indicating higher intensities.

This plot helps in visualizing the relationship between the soliton propagation and the third mover's strategy, showing how the localized wave packet (soliton) moves and spreads over time, representing the evolution of the third mover's strategy under the plasmon resonance conditions.

This 3D visualization illustrates how the distribution of the Third Mover's strategy, modeled by the soliton solution $S_C(x,t) = A_C \text{sech}^2(B_C(x - C_C t))$, evolves over time and space. When considering this distribution from the perspective of fake news diffusion risk and digital health, the following insights may be gained:

The soliton solution exhibits a localized waveform that moves and spreads over time. In the context of fake news, this waveform can represent the strength and impact of the disseminated information. As indicated by this model, the influence of fake news may be dramatically amplified under specific conditions (e.g., plasmon resonance conditions). By adjusting strategies based on information from preceding sources (First Mover and Second Mover), the Third Mover may further amplify the diffusion of information.

This 3D model provides a useful visual representation for understanding the dynamics of information diffusion and devising strategies to mitigate the risks of fake news. By employing such models, insights can be gained for developing practical approaches to counteract fake news and formulate policies and educational programs to promote digital health.

# 18. Discussion:Third mover,Setting up Nonlinear Partial Differential Equations

To model the strategy $S_C(x, t)$ of the third mover, we consider the following nonlinear partial differential equation:

$$\frac{\partial S_C}{\partial t} + \alpha_C S_C \frac{\partial S_C}{\partial x} = \beta_C \frac{\partial^2 S_C}{\partial x^2} + \kappa_C S_C^2 + G(S_C, x, t, S_A, S_B)$$

Here, $\alpha_C$ represents the coefficient of self-enhancement for the third mover, $\beta_C$ represents the diffusion effect of the third mover, $\kappa_C$ represents the coefficient of the plasmon resonance condition, indicating the amplification of diffusion under certain conditions, $G(S_C, x, t, S_A, S_B)$ represents the strategic response of the third mover based on the strategies of the first and second movers.

We assume the following ansatz for the soliton solution:

$$S(x, t) = A \operatorname{sech}^2(B(x - Ct))$$

Time Derivative:

$$\frac{\partial S}{\partial t} = -2ABC \operatorname{sech}^2(B(x - Ct)) \tanh(B(x - Ct))$$

Spatial Derivative:

$$\frac{\partial S}{\partial x} = 2AB \operatorname{sech}^2(B(x - Ct)) \tanh(B(x - Ct))$$

$$\frac{\partial^2 S}{\partial x^2} = 2AB^2 \operatorname{sech}^2(B(x - Ct))[2\tanh^2(B(x - Ct)) - 1]$$

We substitute the above differentiation results into the original nonlinear partial differential equation and simplify each term to obtain the following expression:
$$-2ABC \operatorname{sech}^2(B(x - Ct)) \tanh(B(x - Ct))$$
$$+ \alpha A \operatorname{sech}^2(B(x - Ct)) \tanh(B(x - Ct))$$
$$= \beta 2AB^2 \operatorname{sech}^2(B(x - Ct))[2\tanh^2(B(x - Ct)) - 1]$$
$$+ \kappa A^2 \operatorname{sech}^4(B(x - Ct))$$

We substitute the above differentiation results into the original nonlinear partial differential equation and simplify each term to derive conditions on the parameters $A_C$, $B_C$, and $C_C$ of the soliton solution.

By comparing each term of the equation, we derive conditions on $A_C$, $B_C$, and $C_C$. In particular, we analyze the effect of the plasmon resonance condition ($\kappa_C A_C^2$ term) on the characteristics of the soliton solution of the third mover.

Considering the effect of the plasmon resonance condition on the soliton solution of the third mover, we analyze how the diffusion of fake news is amplified. In particular, we investigate how the amplitude $A_C$ of the soliton solution changes as the value of $\kappa_C$ increases.

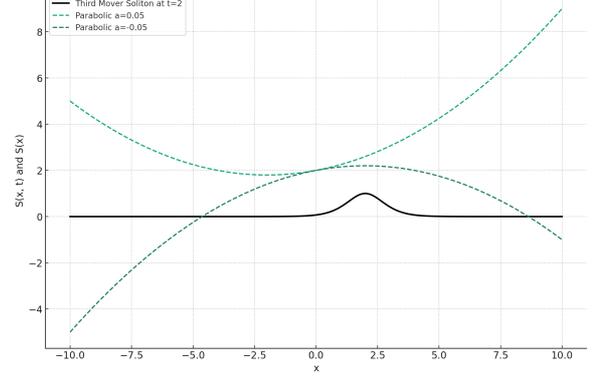

Fig. 20: Comparison of Third Mover Soliton Solution and Parabolic Strategies under Plasmon Influence

To visualize the relationship between the Third Mover, Plasmon, and the parabolic condition, we plot the distribution representing the Third Mover's soliton solution $S_C(x, t) = A_C \operatorname{sech}^2(B_C(x - C_C t))$ and the parabolic function $S(x) = ax^2 + bx + c$ on the same graph. This allows us to compare the shape of the Third Mover's strategy at a specific time with the shape defined by the parabolic condition. The influence of Plasmon is considered in the shape and propagation of the soliton solution.

We plot the soliton solution at a constant time $t$, overlaid with different parabolic functions on the same x-axis. This visual representation enables us to compare how the propagation shape of the Third Mover's strategy relates to the shape defined by the parabolic condition.

The graph above compares the Third Mover's soliton solution $S_C(x, t) = A_C \operatorname{sech}^2(B_C(x - C_C t))$ at a specific time $t = 2$ with parabolic functions $S(x) = ax^2 + bx + c$ with different $a$ values. The soliton solution is depicted by the solid black line, while the parabolic strategies are represented by dashed lines.

This visualization helps understand how the Third Mover's strategy evolves over time and how both Plasmon's influence and the parabolic condition affect it. The differences in shape between the soliton solution and the parabolic strategies indicate considerations when modeling different strategic behaviors or physical phenomena.

This graph compares the soliton solution of the Third Mover at a specific time with parabolic strategies. The soliton solution represents the strategy of the Third Mover influenced by Plasmon, while the parabolic strategy demonstrates how the curve of the strategy changes with different values of $a$.

The soliton solution may represent the strength and speed of fake news diffusion. The shape of this localized wave indicates rapid spread of fake news followed by decay over time.

The curve of the parabolic strategy may represent individual or collective responses to fake news. Positive values of *a* indicate increasing reactions of people as fake news spreads, while negative values indicate initially strong reactions followed by a decrease in response.

To minimize the risk of fake news, strategies that reduce the peak of the soliton solution (i.e., decreasing $A_C$) are necessary. Additionally, adjusting the curve of the parabolic strategy is important to appropriately guide societal responses to fake news.

The propagation of strategies by the soliton solution may represent something akin to herd immunity against fake news, reflecting strategies to minimize the psychological stress and societal unrest caused by the spread of fake news.

The parabolic strategy may model the impact of societal measures such as education and awareness campaigns against fake news, potentially altering individual perceptions and behaviors towards fake news.

Enhancing digital literacy to effectively identify and address fake news is essential for maintaining digital health.

Such graphs serve as valuable tools for providing insights into the risks of fake news diffusion and digital health. Using these models, understanding which strategies are most effective when formulating responses to fake news can be achieved.

## 19. Perspect

In this study, we have made a new attempt to model the process of fake news diffusion using incomplete information games and nonlinear partial differential equations. In particular, we theoretically analyzed the phenomenon of the rapid amplification of fake news under certain social conditions, analogous to plasmon resonance in physics. In addition, we used game theory concepts to explore how first mover, second mover, and third mover strategies affect the dynamics of fake news diffusion.

This approach allowed us to show that the spread of fake news is not simply the transmission of information, but the result of complex strategic interactions. In addition, the amplification effect of fake news under plasmon resonance conditions provides an important perspective for understanding how information is more likely to spread in specific social groups and communication contexts.

The approach of this study involves several challenges. The proposed model is based on a theoretical framework that needs to be further tested for its applicability to actual fake news cases and data. The challenge is to define parameters related to plasmon resonance conditions and players' strategies in line with real situations. 2. The analysis of nonlinear partial differential equation models is complex and requires further validation through numerical simulations. It is also important to evaluate the validity of the model by comparing its predictions with actual patterns of fake news dissemination. The spread of fake news is not simply a matter of information transmission, but is influenced by diverse factors such as human cognitive biases, social influences, and cultural backgrounds. Incorporating these factors into the model will allow for a more realistic analysis, but the challenge is to develop the theoretical and methodological framework to do so.

In order to understand the mechanisms of fake news proliferation and to provide theoretical insights for its prevention and countermeasures, the above-mentioned challenges need to be addressed. Future research will require numerical simulations based on broader data sets, experimental validation, and the development of new theoretical frameworks to integrate concepts from the social sciences and physics. This will hopefully lead to the development of new strategies to control the spread of fake news and promote informed and fair social debate.

The introduction of plasmon resonance and soliton solutions has several advantages in analyzing the spread of fake news, but also comes with challenges. The concept of plasmon resonance can be used to theoretically capture how fake news can be dramatically amplified under certain conditions. This metaphorical application allows for a deeper understanding of information diffusion mechanisms.

Using the soliton solution, the diffusion of fake news can also be modeled as a nonlinear partial differential equation. This approach explicitly expresses the nonlinear dynamics of information diffusion and captures the complexity of the diffusion process.

By combining the concepts of first mover, second mover, and third mover, we can analyze how different players' strategies affect the diffusion of fake news. Using the soliton solution framework, it is possible to mathematically represent these strategic interactions and track the evolution of strategies.

The challenge is that although the model using plasmon resonance and soliton solutions is based on a theoretical framework, it needs to be tested to see how well it fits actual fake news dissemination data. Validation of the model by comparison with real data is an issue.

The diffusion of fake news is deeply influenced by social, cultural, and psychological factors. The challenge is to what extent mathematical models of plasmon resonance and soliton solutions can handle these complex factors.

Parameterization and interpretation in models of plasmon resonance and soliton solutions can be counterintuitive. In particular, the social scientific interpretation of the parameter $\kappa$, which represents the plasmon resonance condition, and the application of the soliton solution parameters *ABC* to the actual diffusion process require further study.

While the introduction of plasmon resonance and soliton solutions provide useful insights in the analysis of fake news diffusion, challenges remain in their application to real data and social scientific interpretation. In future research, it is important to apply these theoretical models to actual social phenomena to test their validity. Future work is also required to develop methodologies to incorporate the complexity of social context and human behavior into the models.